\documentclass[conference]{IEEEtran}

\usepackage{cite}
\usepackage{amsmath,amssymb,amsfonts}
\usepackage{graphicx}
\usepackage{textcomp}
\usepackage{xcolor}
\usepackage{microtype}
\usepackage{subcaption}
\usepackage{xspace}
\usepackage{hyperref}
\usepackage[capitalise]{cleveref}
\usepackage{booktabs}
\usepackage{array}
\usepackage{placeins}
\usepackage[ruled,linesnumbered]{algorithm2e}
\usepackage{paralist}
\usepackage{scratch3}
\usepackage{tcolorbox}
\usepackage{listings}
\usepackage{xcolor}
\usepackage{balance}
\usepackage[%
  group-minimum-digits=4,%
  list-final-separator={, and },%
  add-integer-zero=false,%
  free-standing-units,%
  round-mode=figures,%
  round-precision=3,%
  binary-units,%
  detect-weight=true,%
  detect-inline-weight=math,%
]{siunitx}
\usepackage{tikz}

\newcommand\definetool[2]{\newcommand{#1}{{\textsc{#2}}\xspace}}

\definetool{\Scratch}{Scratch}
\definetool{\Whisker}{Whisker}
   
\graphicspath{{images/}}

\definecolor{codegreen}{rgb}{0,0.6,0}
\definecolor{codegray}{rgb}{0.5,0.5,0.5}
\definecolor{codepurple}{rgb}{0.58,0,0.82}
\definecolor{orange}{rgb}{0.80,0.40,0.00}

\lstdefinelanguage{JavaScript}{
	keywords={await, break, case, catch, continue, const, let, debugger, default, delete, do, else, finally, for, function, if, in, instanceof, new, return, switch, this, throw, try, typeof, var, void, while, with},
	morekeywords={class, export, boolean, throw, implements, import, this},
	comment=[l]{//},
	morecomment=[s]{/*}{*/},
	string=[b]",
	morestring=[b]',
	sensitive=true
}

\lstdefinestyle{mystyle}{
	commentstyle=\color{codegreen},
	keywordstyle=\color{orange},
	numberstyle=\tiny\color{codegray},
	stringstyle=\color{codepurple},
	basicstyle=\ttfamily\footnotesize,
	breakatwhitespace=false,         
	breaklines=true,                 
	captionpos=b,                    
	keepspaces=true,                 
	numbers=left,                    
	numbersep=5pt,                  
	showspaces=false,                
	showstringspaces=false,
	showtabs=false,                  
	tabsize=2
}
\begin{document}

\title{Model-based Testing of Scratch Programs}

\author{
\IEEEauthorblockN{Katharina Götz}
\IEEEauthorblockA{University of Passau\\
Passau, Germany}
\and
\IEEEauthorblockN{Patric Feldmeier}
\IEEEauthorblockA{University of Passau\\
Passau, Germany}
\and
\IEEEauthorblockN{Gordon Fraser}
\IEEEauthorblockA{University of Passau\\
Passau, Germany}
}

\maketitle

\begin{abstract}
	Learners are often introduced to programming via dedicated languages such
as \Scratch, where block-based commands are assembled visually in order to
control the interactions of graphical sprites.
	Automated testing of such programs is an important prerequisite for
supporting debugging, providing hints, or assessing learning outcomes. However,
writing tests for \Scratch programs can be challenging: The game-like and
randomised nature of typical \Scratch programs makes it difficult to identify
specific timed input sequences used to control the programs. Furthermore,
precise test assertions to check the resulting program states are incompatible
with the fundamental principle of creative freedom in programming in \Scratch,
where correct program behaviour may be implemented with deviations in the
graphical appearance or timing of the program.
The event-driven and actor-oriented nature of \Scratch programs, however,
makes them a natural fit for describing program behaviour using finite state
machines. In this paper, we introduce a model-based testing approach by extending \Whisker, an automated testing framework for \Scratch programs.
The model-based extension describes expected program behaviour
in terms of state machines, which makes it feasible to check the abstract behaviour of a program
independent of exact timing and pixel-precise graphical details, and to automatically derive test inputs testing even challenging programs.
A video demonstrating model-based testing with \Whisker is available at the following URL:
\begin{center}
	\url{https://youtu.be/edgCNbGSGEY}
\end{center}

\end{abstract}


\section{Introduction}

With more than 85 million programs created and published by young learners\footnote{[October 2021] https://scratch.mit.edu/statistics/},
\Scratch~\cite{Scratch} represents the most popular block-based programming language.
Languages like \Scratch support learners by allowing them to drag and drop
programming statements to visually arrange programs without having to remember programming
syntax. As a consequence, all programs are immediately syntactically correct,
but they may still be semantically incorrect.
This creates a need for automated testing tools that help learners implement, debug, and test their programs by providing valuable feedback.

\begin{figure}[tb]
	\centering
	\begin{subfigure}{0.45\columnwidth}
		\includegraphics[width=\textwidth]{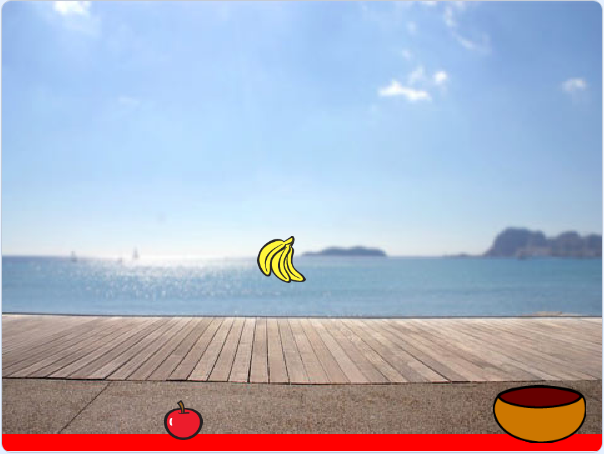}
		\caption{Fruit catching game.}
		\label{fig:OriginalFruitcatcher}
	\end{subfigure}
	\hfill
	\begin{subfigure}{0.45\columnwidth}
		\includegraphics[width=\textwidth]{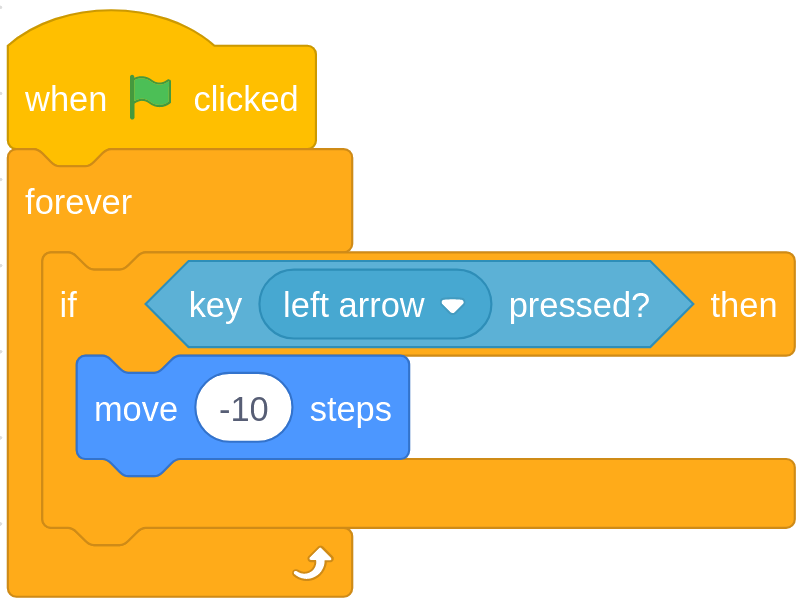}
		\caption{Part of the \Scratch code for bowl sprite.}
		\label{fig:CodeBowlLeft}
	\end{subfigure}\\
	\begin{subfigure}{\columnwidth}
	\begin{lstlisting}[language=JavaScript,numbers=none]
		const test = async function(t) {
			let sprite = t.getSprite('Bowl');
			let oldX = sprite.x;
			t.inputImmediate({device: 'keyboard', 
			         key: 'Left', isDown: true});
			await t.runForTime(1000);
			t.assert.ok(oldX >= sprite.x);
			t.end();
		}
	\end{lstlisting}\vspace{-2em}
	\caption{\label{code:BowlLeft_dumb}\Whisker test for bowl left movement.}
	\end{subfigure}
	\begin{subfigure}{\columnwidth}
		\centering
		\includegraphics[width=0.55\columnwidth]{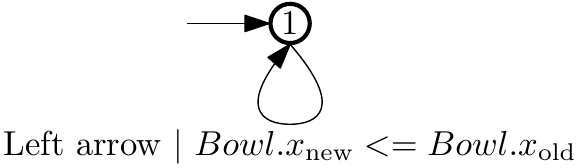}
		\caption{Model testing the left movement of the bowl.}
		\label{fig:modelLeftBowl}
	\end{subfigure}
	\caption{Fruit catching game in Scratch.}
\end{figure}

\Cref{fig:OriginalFruitcatcher} shows an example \Scratch program, in which a
player can move a bowl sprite horizontally using the left and right arrow keys, with the goal of
catching as many fruit sprites falling from the top as possible. \cref{fig:CodeBowlLeft}
shows a code excerpt that handles part of the bowl's behaviour by first checking in
an infinite loop whether the left key was pressed, and then moving the bowl to
the left if the check passes. This behaviour can be examined using the automated
\Whisker~\cite{Whisker} test shown in \cref{code:BowlLeft_dumb}.
The test retrieves the bowl sprite, stores the current x-position, then presses
the left cursor key for 1000ms, waits, and finally checks whether the bowl is now to
the left of where it was previously. While this is an intuitive and easy test,
there are several challenges:
\begin{itemize}
	\item  Being able to write such a test requires
knowledge of JavaScript, which may be problematic if the user writing tests is
a school teacher without thorough software engineering training. 
	\item  The
bowl may make arbitrary wrong movements that will not be detected as long
as the final position after 1000ms is somewhere to the left. 
	\item  It is challenging to
estimate a precise waiting duration before checking the resulting behaviour, as
differences in the computer as well as implementation choices may lead to
different execution times. For example, the duration of an animation will depend on the current sprite position as well as its size and shape.
	\item  A further problem not shown by this example is that test assertions on absolute positions are difficult, as it is common and accepted behaviour in \Scratch for young learners to use arbitrary sprites, shapes, and sizes.
\end{itemize}

To address these issues, in this paper we introduce the notion of model-based
testing for \Scratch. The behaviour of the bowl sprite in
\cref{fig:CodeBowlLeft} can easily be described using the extended finite state
machine shown in \cref{fig:modelLeftBowl}: There is a single state which is
entered when the program is started, and then whenever the left cursor key is
pressed a transition is triggered in which the expected effect is that the
x-position of the bowl has changed to the left. This model can immediately
serve to check the behaviour of the program during execution in conjunction
with any automated tests, or even automated test
generation~\cite{deiner2020Scratch}. While it is unlikely that an
automated test generator will produce sequences of events that will properly
play the fruit catching game, it is possible to also specify the input
behaviour using finite state machines.

In detail, the contributions of this paper are as follows:
\begin{itemize}
	\item We introduce the notion of model-based testing for \Scratch.
	\item We extend the \Whisker testing framework for \Scratch with an editor to create models for testing \Scratch programs.
	\item We provide a fully automated testing solution for \Scratch programs by allowing users not only to model the desired program behaviour, but to also derive test inputs from user models.
	\item We demonstrate the feasibility of the approach using the fruit catching game as a case study and apply model-based testing to 38 students' implementations of the game.
\end{itemize}

\section{Background}\label{Background}

\Scratch~\cite{Scratch} is a popular block-based programming language intended
to support learners new to the world of programming. Programs are created by
dragging and dropping blocks from drawers containing all available blocks, such
that learners do not have to memorise all available commands initially. Furthermore, since
blocks have different shapes, only syntactically valid combinations
are possible, thus avoiding syntax errors. Programs created this way consist of
multiple scripts that define the behaviour of sprites interacting on top of a
graphical stage. The resulting programs can usually be controlled by user
inputs, and high-level program statements make it possible to quickly and
easily create fun and engaging programs and games.

While syntax errors are prevented by \Scratch, learners may still struggle to
produce functionally correct programs, and teachers may need help in supporting
their students. Therefore, automated tools are an important means for
supporting both of these groups. For example, there are tools that report code
smells~\cite{hermans2016smells,boe2013hairball,moreno2015dr,techapalokul2017quality,chang2018scratch} or bug patterns~\cite{fradrich2020common}, but such tools
are limited to generic feedback independent of a task at hand. In order to
produce feedback on whether a specific target functionality is satisfied, or to
generate hints on how to proceed to get there, automated tests are an important
prerequisite~\cite{keuning2016towards}. 

Testing interactive, graphical programs like those typically created with
\Scratch comes with multiple challenges, such as the heavily randomised nature
of game-like programs, long execution times caused by animations and story-like
segments of programs, or their event-driven and graphical nature. The
\textsc{Itch}~\cite{johnson2016itch} tool attempts to provide testing functionality by
translating a subset of the language features to Python, but is restricted to
checking textual dialogue behaviour. We have therefore introduced the \Whisker
testing framework~\cite{Whisker} which makes it possible to write automated
tests for \Scratch programs using a JavaScript API as shown in
\cref{code:BowlLeft_dumb}. To support users, \Whisker can also automatically
generate test inputs (e.g. randomly or using search~\cite{deiner2020Scratch}),
but in order to detect faults there nevertheless has to be a specification to
check against, which so far had to be written in JavaScript. Both, writing
tests as well as the specifications, are non-trivial tasks, in particular
considering that an important target audience are users who may
only have limited software engineering training (e.g. teachers or creators of
\Scratch tutorials).

In many software engineering domains, model-based approaches have been
successfully applied to improve testing~\cite{dias2007survey}. Among the many
notations available to specify system behaviour~\cite{utting2012taxonomy},
variants of finite-state machine notations are among the most common and longest
established ones~\cite{lee1996principles}. Finite-state models have also been a
frequent subject of research for automatically deriving
tests~\cite{anand2013orchestrated}. However, to the best of our knowledge,
neither the use of model-based testing in general, nor the use of finite-state
machine notation specifically, have been explored in the context of block-based
programming languages like \Scratch yet.


\section{Model-based Testing for Scratch Programs}\label{sec:ModelBasedTesting}

\subsection{Modelling Scratch Programs}

We aim to model the state-based behaviour of \Scratch programs as finite state
machines. The concrete state of a \Scratch program is defined by the values
assigned to variables and attributes of sprites contained within a program
(e.g. position, rotation, size, visibility, graphics effect, costume).
To keep the size of models small, we use \emph{abstract} states, where an
abstract state (1) refers to a subset of the variables and attributes of a
\Scratch program, and (2) can represent multiple concrete states of the
remaining variables and attributes. This means that there can be multiple
models for a single \Scratch program, each representing different aspects of
behaviour.

We define a model for a \Scratch program as an extended finite
state machine ($S, q_0, Q_0, Q_1, \Sigma, E)$, where:
\begin{itemize}
	\item $S$ denotes a finite set of
abstract \emph{states}.
\item $q_0\in S$ is the \emph{initial state} of the machine.
\item  $Q_0\subset S$
specifies a set of \emph{stop states} that halt the execution of the corresponding model.
\item $Q_1 \subset Q_0$ defines an
additional set of \emph{stop states}, where each node stops the execution of \emph{all} models for the given \Scratch program. 
%
\item $\Sigma: P \rightarrow \{0, 1\}$ defines a set of \emph{predicates} that may hold for a given concrete program state $P$. As the \Scratch language is event-driven and action-oriented, this set consists of predicates describing the presence or absence of user inputs like key presses and mouse clicks, and of graphical program events such as two sprites touching each other. Additionally, we include time events such as `two seconds elapsed'. 
%
%
%
\item $E: S \times 2^\Sigma \rightarrow S \times 2^\Sigma$ is a finite set of state transitions, where $t(s, C): (s', C')$ denotes that from state $s \in S$ given all of $c \in C \subset \Sigma$ evaluate to true, we move in a step to $s' \in S$ with all $c' \in C' \subset \Sigma$ evaluating to true. The \emph{state transitions} $E$ enforce chronological dependencies between all \emph{conditions} $c\in C$ and between all \emph{effects} $c'\in C'$.
%
%

\end{itemize}

\begin{figure}[tb]
	\centering
	\includegraphics[width=0.45\columnwidth]{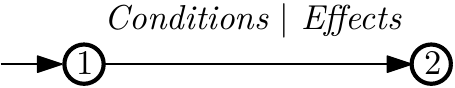}
	\caption{\label{fig:notation}Notation for state machines.}
\end{figure}

We represent models using the notation shown in \cref{fig:notation}: States are
represented as circles, and a transition $t(1, \text{Conditions}): (2,
\text{Effects})$ is represented as a directed edge from state $1$ to state $2$,
where we separate conditions and effects using the $|$ symbol. For example, the model corresponding to the bowl sprite
shown in \cref{fig:modelLeftBowl} has one state with a single
self transition bound to the condition `left arrow was pressed'. If the condition evaluates to true, the program is tested
against the effect of the bowl moving to the left.

A \emph{program model} is a model of a \Scratch program where the
pseudo-transition to the initial state is activated by the greenflag-event,
which represents the user initiating the program execution by pressing the
greenflag button in \Scratch. Program models describe the behaviour 
until the program execution is stopped, that is when the specification defines an abort
or a valid end, represented by the stop states.

In order to simplify the description of various program behaviours, we further introduce an alternative to program models called
\emph{end models}, in which the pseudo-transition to the
initial state is activated as soon as all program models have halted.
For the testing process, end models are optional, but they are useful for
testing conditions that should hold after any stopping condition in the
specification is reached, such as not changing the player's points anymore. For program
and end models, we restrict the global stop states $Q_1$ to only halt state
machines of the same type.


\Scratch models can be executed in parallel to both, manually written \Whisker tests as well as tests automatically generated by one of \Whisker's test-generation algorithms.
Additionally, we introduce \emph{user models} that are capable of specifying possible user inputs as finite state machines.
 More specifically, a user model $U = (S, q_0, Q_0,
Q_1,\Sigma, \Sigma_I, E)$ is an extended finite state machine which differs
from a program model in two ways:
\begin{itemize}
	\item $\Sigma_I$ is the set of possible user inputs, such as key presses, mouse movement, mouse clicks.
	\item $E: S \times 2^\Sigma \rightarrow S \times 2^{\Sigma_I}$ defines the transition relation; unlike program models the effects of an edge represent sets of user events from $\Sigma_I$ rather than predicates.
\end{itemize}
When executing a user model $U$ together with a \Scratch program under test, whenever an edge of $U$ is traversed, the set of user inputs that describe the effects of that transition is applied to the program under test by \Whisker.
Consequently, by applying a user model and a set of program models the process of testing \Scratch programs can be automated. By providing conditions that are taken with a defined probability also non-deterministic user models can be built.

\subsection{Testing Scratch Programs with Models}

In order to use models for testing \Scratch programs we execute the program under test
and the models in parallel. \Whisker starts the model test by simultaneously sending a greenflag event to the project under test and setting all program models to their initial states. The parallel execution of the models and the program is achieved by interleaving program execution steps with model updates.

\begin{figure}[tb]
	\centering
	\includegraphics[width=\columnwidth]{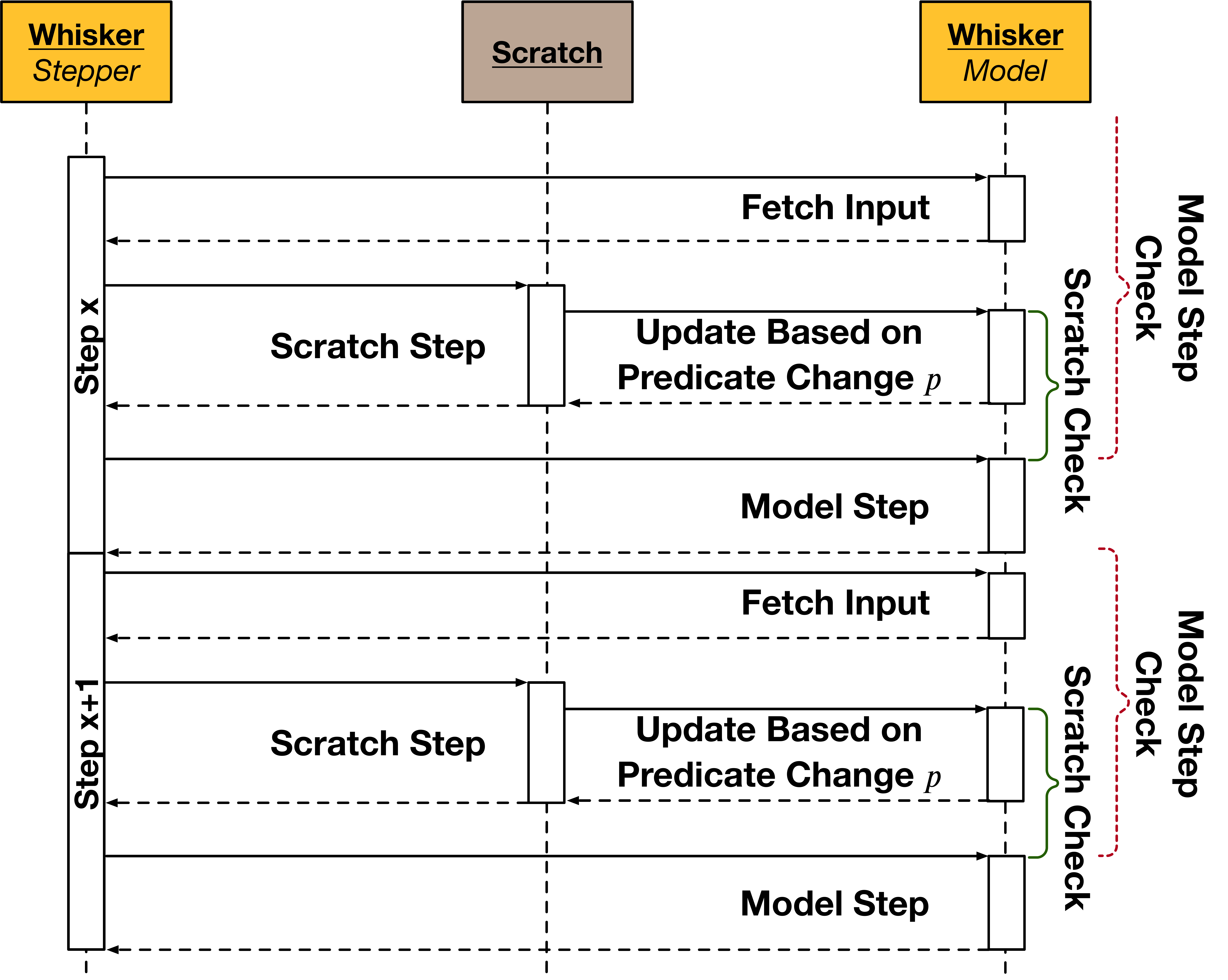}
	\caption{\label{fig:whiskerStep}Model-based \Scratch testing embedded within the \Whisker step function.}
\end{figure}

%
The execution of \Scratch programs is based on a
step function that is invoked at a regular interval, executing all
active scripts (each active script represents one execution thread). The
\Whisker testing framework wraps this step function in order to apply test
inputs and check properties~\cite{Whisker}. \cref{fig:whiskerStep}
illustrates the abstract procedure of a \Whisker step. 

Model-based testing extends the workflow by inserting a model input phase
before and a model step after the \Scratch step.
%
User models generating inputs are triggered in the model input phase to provide the input for the model steps as well as for the \Scratch~Virtual machine (VM).
In the model step after a \Scratch step we update the states of
all active models using the current program state and elapsed time. 

In addition, we instrumented the \Scratch VM such that each time a
predicate $p\in \Sigma$ changes its value \Whisker is informed, and all active model states are updated based on their transition relation $E$. Note that
every unique state transition $e\in E$ can only be triggered once per \Whisker
step in order to traverse loops in the model only a single time.


A \Scratch model may have multiple transitions enabled in the same state. To
avoid non-determinism, the transitions are implemented as an ordered list in
\Whisker, implying a priority on the state transitions. The user can specify
the order for each state when defining the models. Additionally, to reduce
errors in the modelling process \Whisker tests all simultaneously tested effects
of every active model for contradicting logic and removes them from the testing
set. Contradicting effects can be literally opposites, i.e., one excludes the
other and vice versa like testing a boolean for true and false at the same
time.  \Whisker also checks for contradicting ranges of results,
e.g. sprite.x~$>0$ and sprite.x~$<-1$.

Whenever a model executes a transition $t(s,C): (s', C')$, we check whether
all predicates $c' \in C'$ are satisfied by the resulting program state. The
\Scratch program's behaviour is heavily dependent on time, and the exact moment
at which an expected effect manifests in the program state will depend on the
computations performed as well as the rendering process of \Scratch.
Consequently, this raises the question of when and how long to apply the check
to the resulting program state. We therefore check the effect $c'$ for an
interval of time depending on when the transition was triggered: When a
predicate change originates from a transition within the \Scratch step, then the effect $c'$ is checked for
the remainder of the current \Scratch execution step. \cref{fig:whiskerStep} displays this as the \emph{Scratch Check}. If a state transition is
executed during the model step, the effect is checked in the \emph{next}
\Scratch step until the following model step, shown in \cref{fig:whiskerStep} as the \emph{Model Step Check}. In general, the time frame for evaluating the effect of a \Scratch step covers at most one
\Scratch step. The only exception are effects
depending on speech bubbles which are checked for two \Scratch steps, as we
have observed that the \Scratch GUI suffers from delays when rendering these. 

Overall, this allows accurate testing of dependencies between conditions
and effects. If any of the predicates are not satisfied after their time frame,
then the model has found a failure in the implementation under test. Once all
models have reached a stop state, all available end models are started. The
test ends when all program and end models have stopped.

\FloatBarrier


\FloatBarrier
\section{The \Whisker testing tool}

\begin{figure}[tb]
	\centering
	\includegraphics[width=\linewidth]{gui/whisker-main}
	\caption{Main \Whisker GUI.}
	\label{fig:gui-main}
\end{figure}

\begin{figure}[tb]
	\centering
	\includegraphics[width=\linewidth]{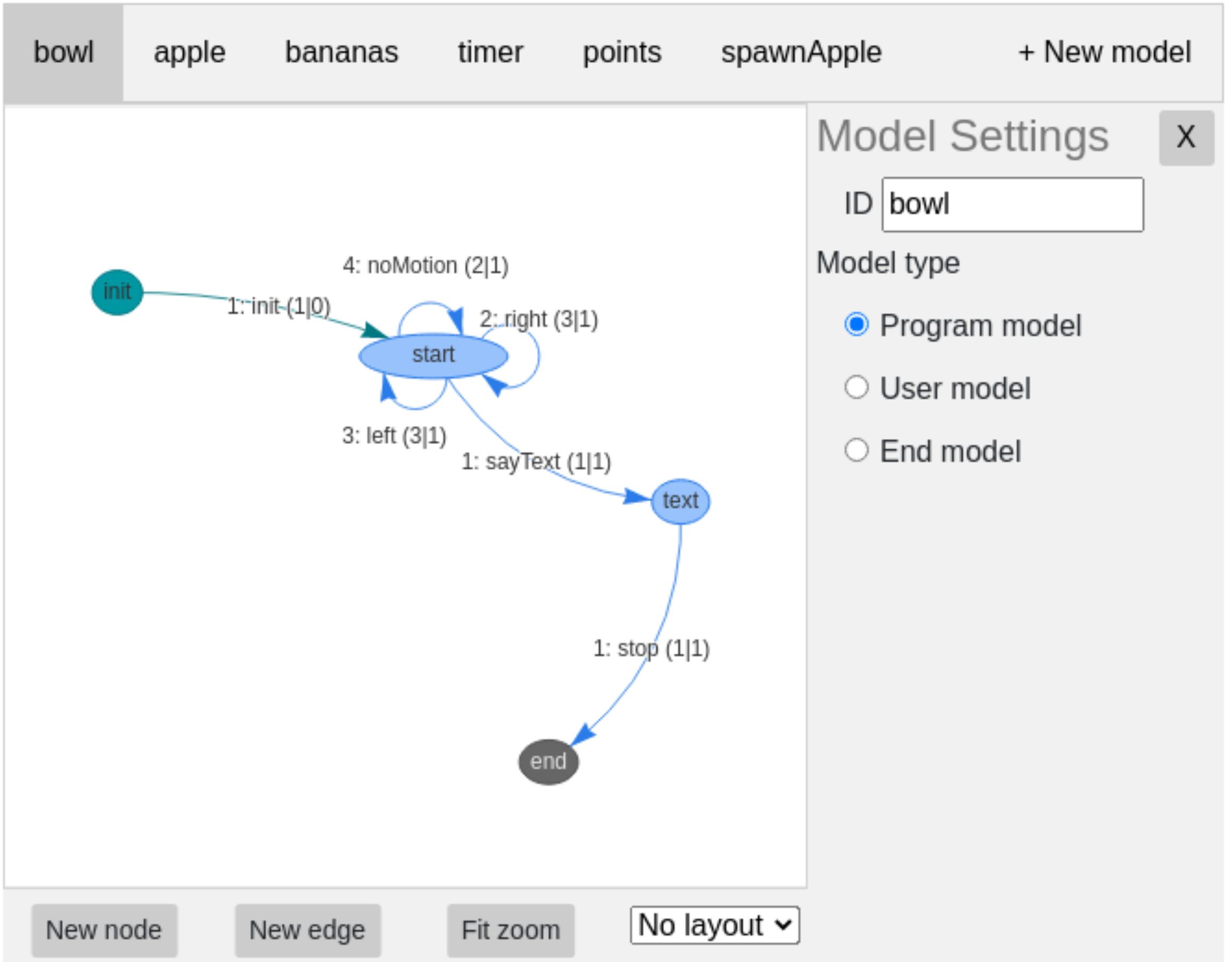}
	\caption{GUI for model-based testing in \Whisker.}
	\label{fig:gui}
\end{figure}

\begin{figure}[tb]
	\centering
	\begin{subfigure}{0.30\columnwidth}
		\includegraphics[width=\textwidth]{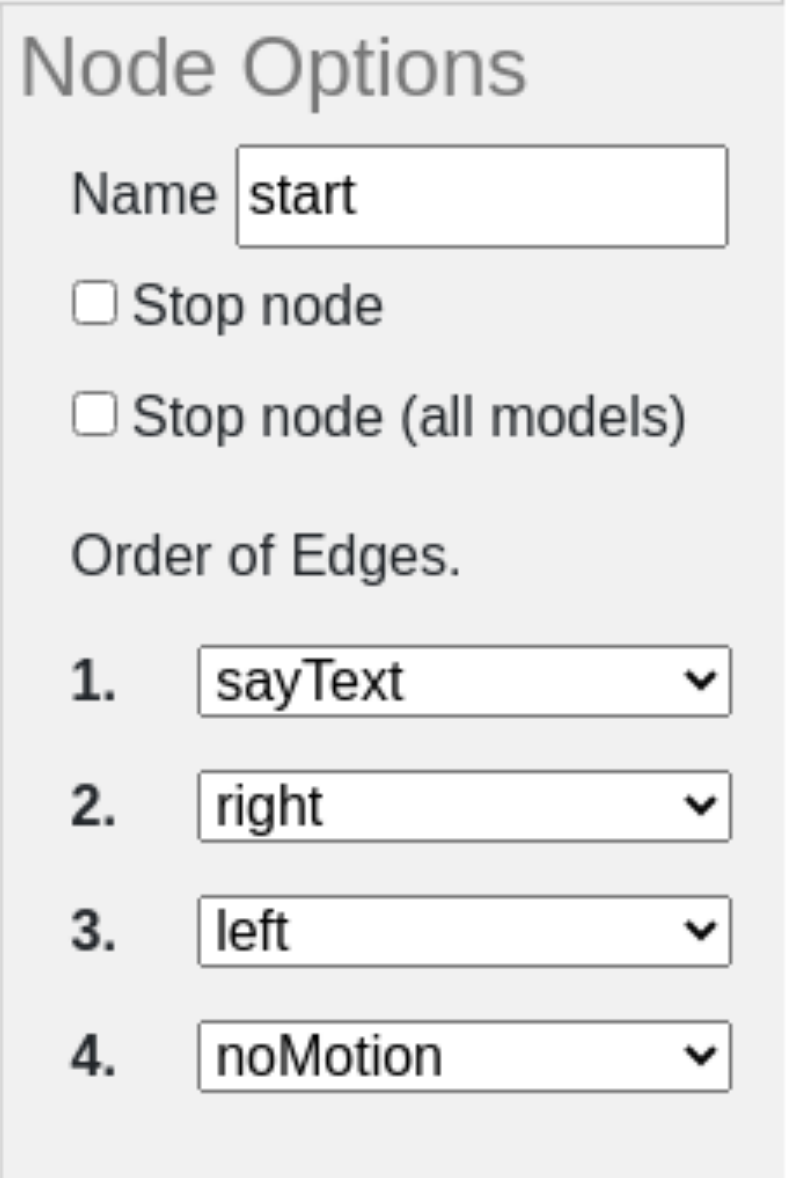}
		\caption{Option panel shown when a node was clicked.}
		\label{fig:nodeOptions}
	\end{subfigure}
	\begin{subfigure}{0.30\columnwidth}
		\includegraphics[width=\textwidth]{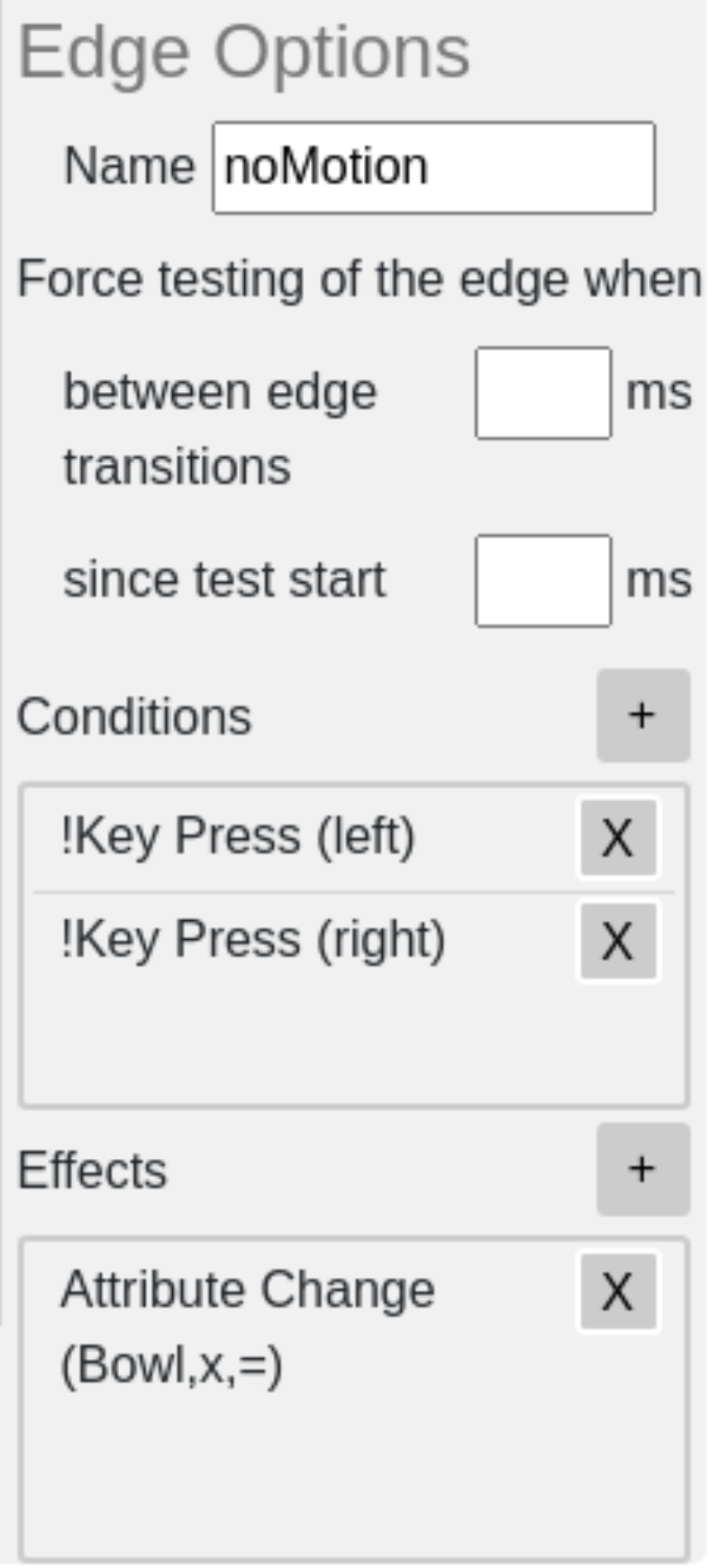}
		\caption{Option panel shown when an edge was clicked.}
		\label{fig:edgeOptions}
	\end{subfigure}
	\begin{subfigure}{0.30\columnwidth}
		\includegraphics[width=\textwidth]{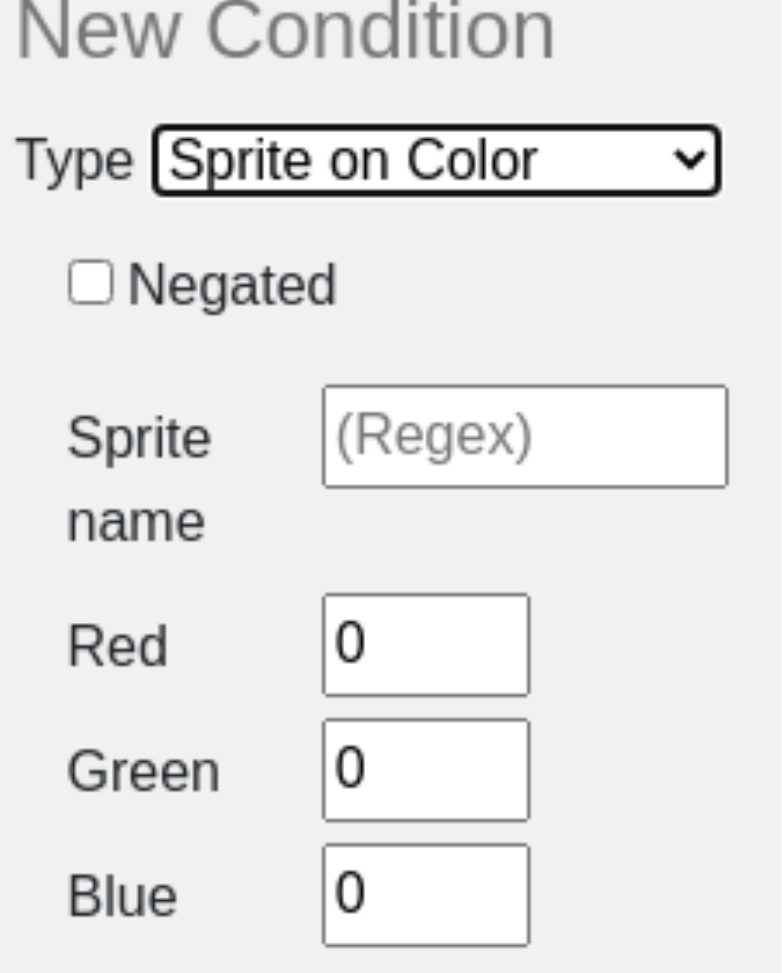}
		\caption{Adding/Editing checks for an edge.}
		\label{fig:newCheck}
	\end{subfigure}\\
	\caption{Based on the user selection, the right panel of the GUI changes to allow configuring the selected part of a model.}
\end{figure}

\Whisker provides a web interface where the user can load and execute a project including corresponding tests (\cref{fig:gui-main}). We extended \Whisker with an intuitive and straightforward editor embedded directly in the web interface, as shown in \cref{fig:gui}.
The tabs at the top can be used for switching between models and adding new ones. The left panel shows an interactive graph, allowing users to modify the position of nodes and change the focus of the view by zooming in and out.
Pressing the \emph{Fit~zoom} button at the bottom resets the zoom and node positions to show the complete graph within the panel.
The editor also supports reordering models in a tree layout, facilitating the editing of smaller as well as bigger models by sorting the graph left-to-right or top-to-bottom beginning at the start node. 
Nodes are labelled and coloured depending on their purpose (e.g. start, stop or stop
all). Edge labels, on the other hand, consist of the order in which the edge transitions should be tested, assigned names, the number
of conditions and the number of effects. Nodes or edges may be added to the graph by using the \emph{New~node} or respectively \emph{New~edge} button on the bottom left of the user interface.
The content of the right panel depends on the user's current selection:
\begin{itemize}
	\item In case nothing is selected, the interface shows the model settings, which allow defining the model name and type.
	\item Clicking on a node
highlights the node including its adjacent edges on the graph and shows the node
option panel instead of the model settings on the right. The node options,
shown in \cref{fig:nodeOptions}, allow changing a node's name, its functionality and
the testing order of the outgoing adjacent edges. Furthermore, to
enforce the presence of exactly one start node in each model, setting a node type to a
start node is not allowed since start nodes are added automatically during the creation of new models.
\item As depicted in \cref{fig:edgeOptions}, selecting an edge offers options for customising the attributes of an edge. The upper half of the panel allows changing the name 
and setting a time constraint on the edge. The time constraint can be defined to either test an edge after a
fixed time passed since the start of the testing process or to be based on the time passed since the last edge transition in the respective model. The lower half shows an overview of the conditions and
effects of the currently chosen edge. 
\item When adding a new check or selecting an existing check, the panel switches to
the check options shown in \cref{fig:newCheck} with the input fields for the arguments shown depending on the chosen
check type. While input fields for checks with fixed values such as keypresses offer predefined values to the user, fields with numeric arguments, such as RGB values, support users with range checks and hints. 
\end{itemize}

\begin{figure}[tb]
	\centering
	\begin{subfigure}{1\columnwidth}
		\includegraphics[width=\textwidth]{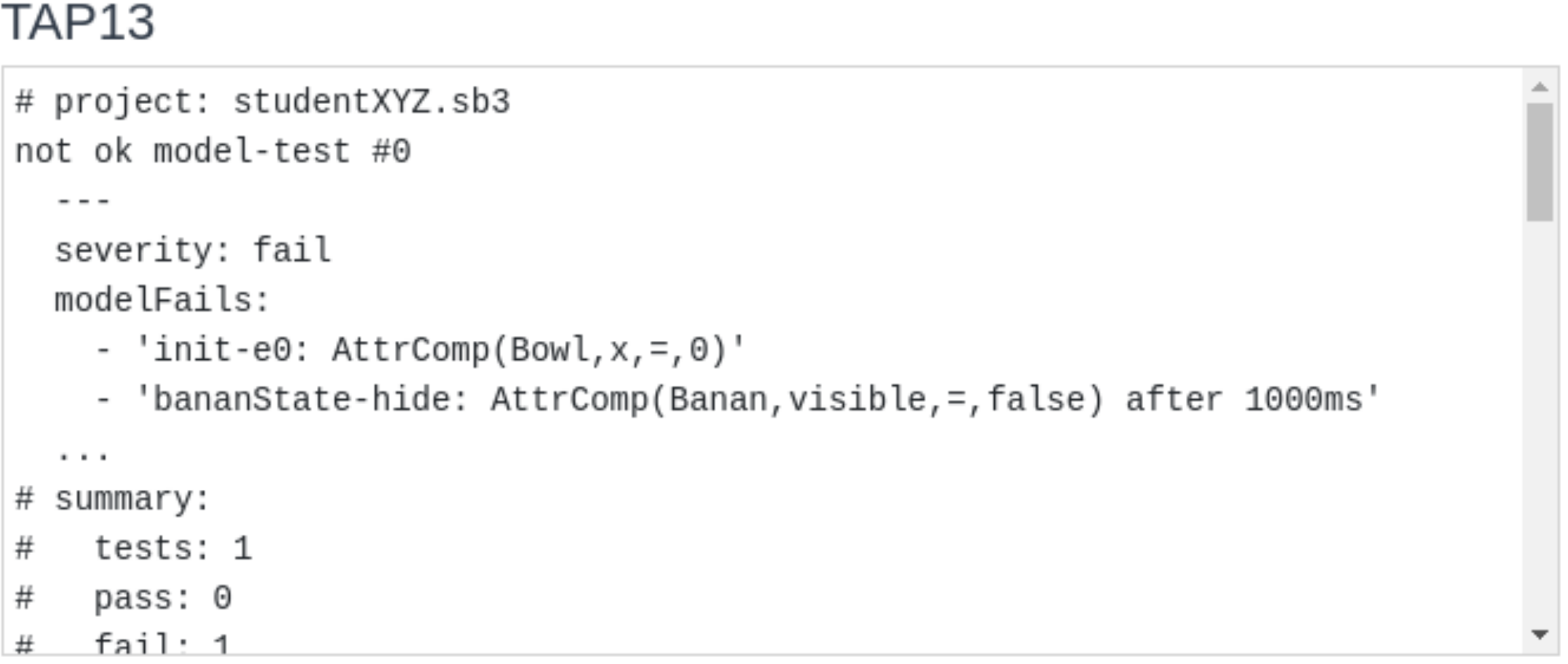}
		\caption{Failures and summary of an executed testing run.}
		\label{fig:tapGui}
	\end{subfigure}\\
	\vspace{0.3cm}
	\begin{subfigure}{1\columnwidth}
		\includegraphics[width=\textwidth]{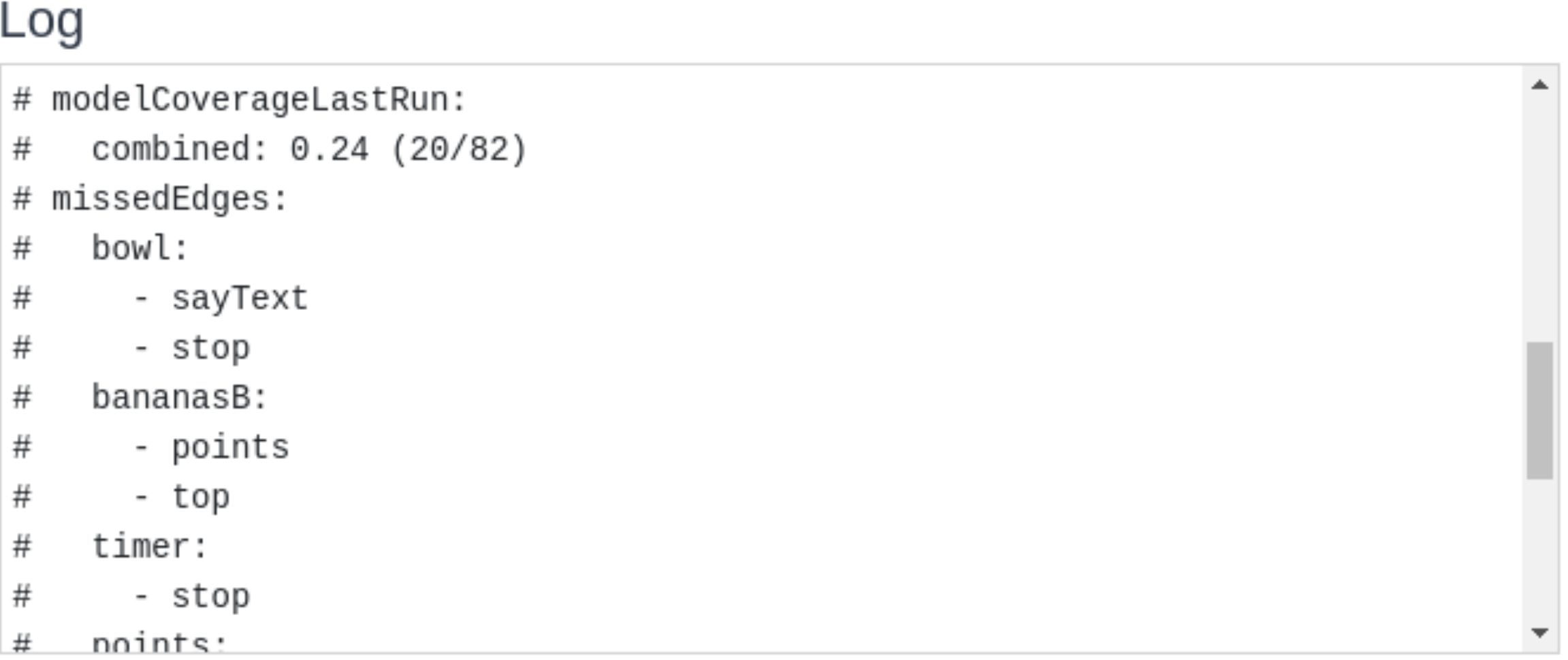}
		\caption{Model details corresponding to an executed testing run.}
		\label{fig:logGui}
	\end{subfigure}
	\caption{The test results including a summary shown by \Whisker at the end of a test.}
\end{figure}

Using the \Whisker interface, users can download created models from the editor as a JSON
file or load existing ones into the testing framework to start the testing process immediately.
These functions offer users the comfortable workflow of loading existing models, modifying them to match their desired program behaviour, and finally saving them again on their system.
When clicking the play button on the left in \cref{fig:gui-main}, the test framework starts testing the currently loaded project file.
Depending on the uploaded files, the program under test can either be tested using a test suite written in JavaScript, models or both.
However, if the user decides to test a given implementation using both testing approaches, he/she has to ensure that only one of these approaches provides user inputs. 
Moreover, the interface allows users to observe the testing process since the game canvas is constantly updated according to the sent user inputs.
The combined results of the JavaScript tests and models are shown with a
different detail level in two textual representations. 
Short excerpts are shown in 
\cref{fig:tapGui} and \cref{fig:logGui}.
\Whisker, including the new model-based testing feature, can also be executed
on the command-line, where it executes regular and/or model-based tests in a
headless browser.

\section{Case Study}
\label{sec:CaseStudy} 

We conducted a case study based on the fruit catching game
shown in \cref{fig:OriginalFruitcatcher}, which originates from \Scratch
lessons for school children. The fruit catching game consists of an apple and a
banana sprite which drop down from the top of the screen, and the aim of the
game is to prevent falling fruit from touching a red bar located at the bottom of the screen by catching the falling fruit with a bowl that can be moved
horizontally using the cursor keys. The game comes with a detailed textual
specification of the desired behaviour, as well as a very detailed test suite
consisting of 27 \Whisker tests~\cite{Whisker}, which makes it well suited for
formalisation with models. In total, we used
19 program models, one end model and one user model with input generation based on randomness to capture all possible aspects of the program.

Names of sprites or variables can be described in our tool by regular
expressions. All references of apple or bananas in the depicted models are simplified
for clarity but actually are given by `$/$(Apple$|$Apfel)$/$' and
`$/$Banan$/$' in the implemented models. Using regular expressions for names and also for the content of
speech bubbles allows for a choosable degree of accuracy when testing with a
wider range of implementations. 
Our tool also offers testing strings with case sensitivity, which is deactivated by default.


\subsection{Bowl Model}

\begin{figure}[tb]
	\centering
	\begin{subfigure}[b]{0.49\columnwidth}
		\centering
		\includegraphics[width=1\linewidth]{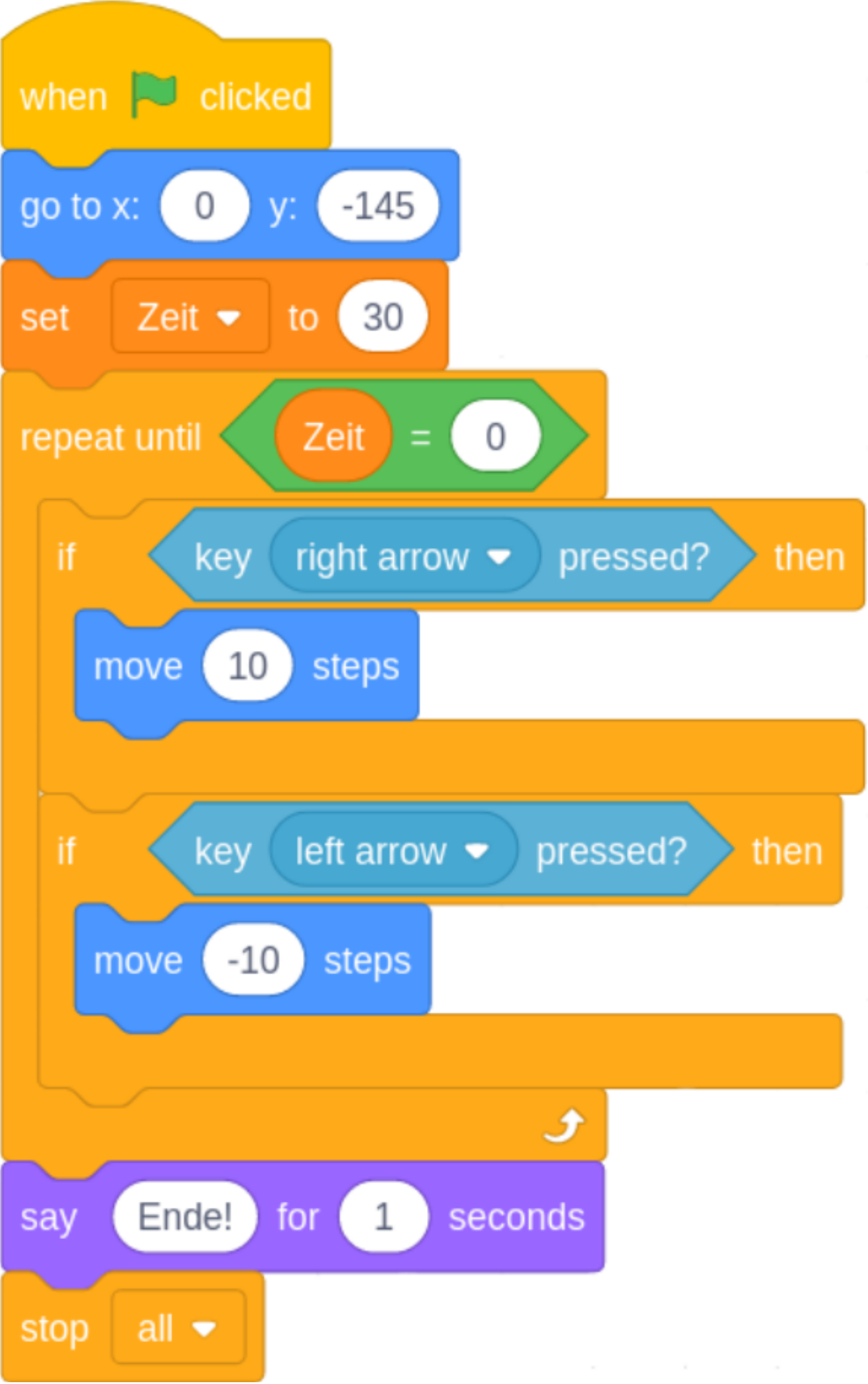}
		\caption{Sample solution for bowl behaviour.}
		\label{fig:codeBowlCorrect}
	\end{subfigure}
	\hfill
	\begin{subfigure}[b]{0.49\columnwidth}
		\centering
		\includegraphics[width=1\linewidth]{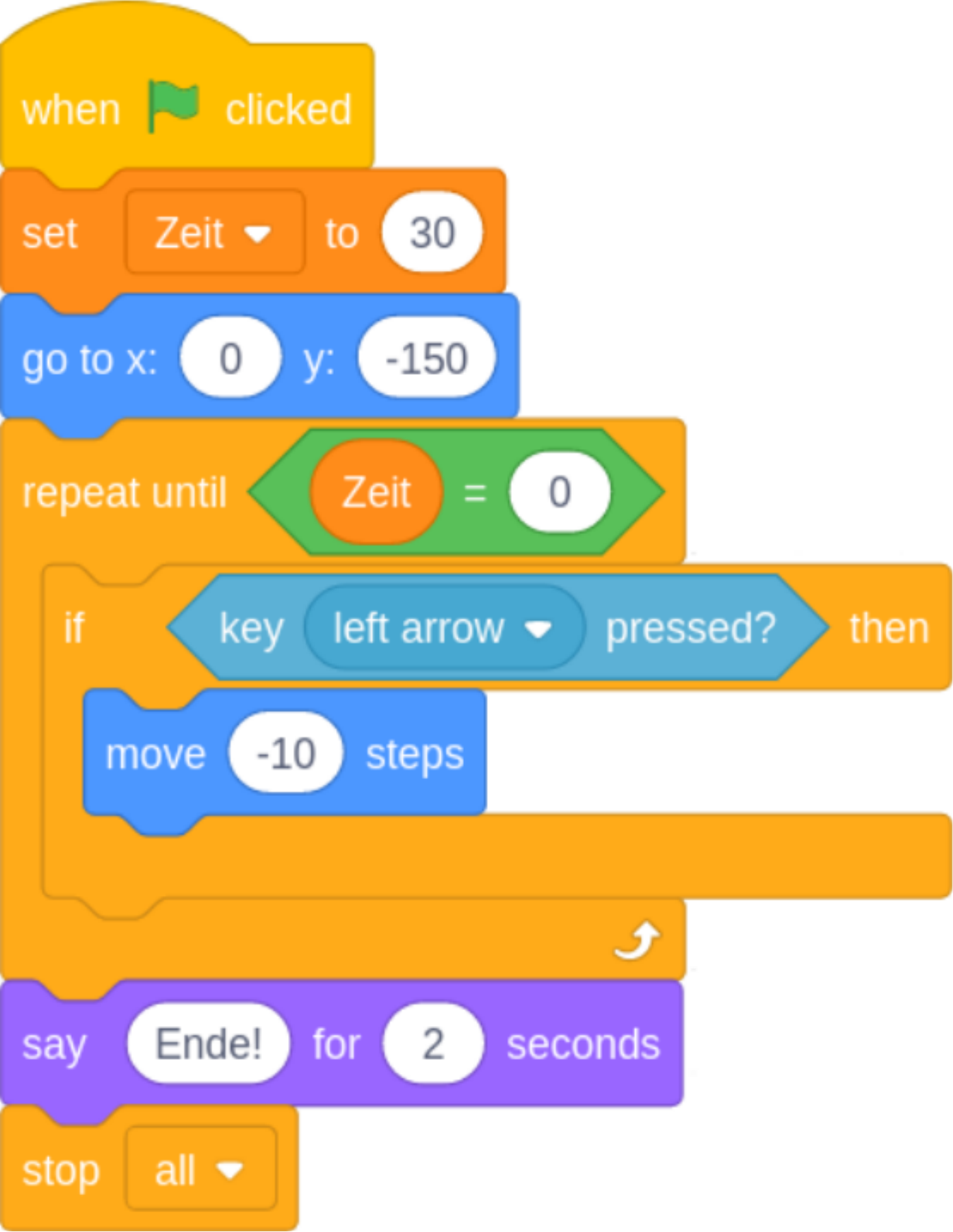}
		\caption{Code example with mistakes for bowl sprite.}
		\label{fig:codeBowlWrong}
	\end{subfigure}
	\caption{Code examples for the bowl sprite.}
	\label{fig:codeBowl}
\end{figure}

\begin{figure}[tb]
	\centering
	\includegraphics[width=\columnwidth]{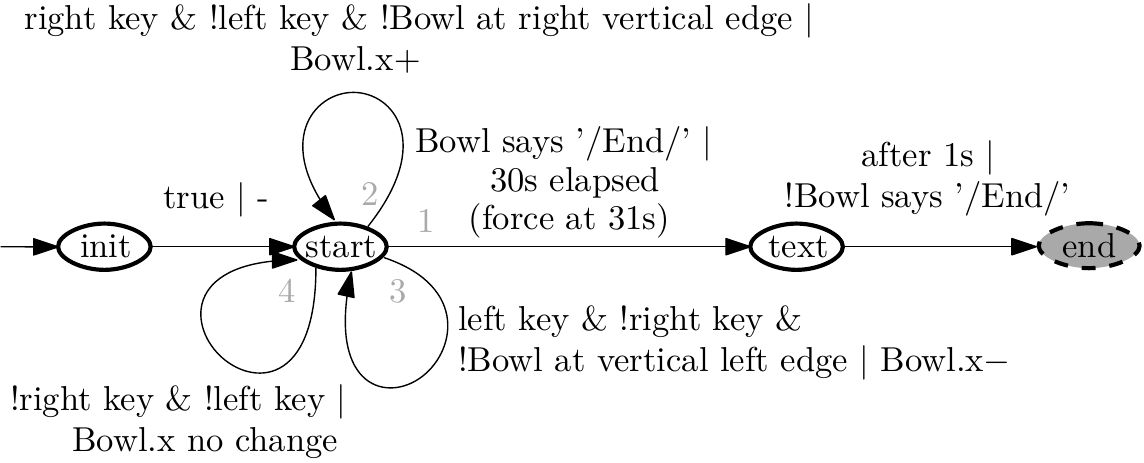}
	\caption{Program model defining the behaviour of the bowl~sprite.}
	\label{fig:BowlProgramModel}
\end{figure}

The behaviour of the bowl sprite consists of two aspects:
\begin{enumerate}
	\item The user can control the bowl sprite by moving it horizontally with
the left and right cursor keys.
	\item The bowl should display a speech bubble when a 30 seconds timer has elapsed.
\end{enumerate}
The expected implementation of the bowl is shown in \cref{fig:codeBowlCorrect},
and \cref{fig:BowlProgramModel} shows the corresponding program model. The
first edge (init, start) is always taken because of its `true' condition and no
effects are checked. We use this initial transition to give the
program time to initialise variables and sprite positions before applying
any checks.

In the `start' state there are four outgoing transitions: Two of them are self-loops that test if the x-position of the bowl increases/decreases in case the cursor keys are pressed and the bowl is not touching a limiting vertical edge of the screen. Another self-loop checks whether the x-position remains unchanged if the right/left cursor keys are not pressed.
The edge (start, text) is triggered as soon as the bowl displays a speech
bubble which text conforms to the regular expression `$/$End$/$' testing that 30 seconds have elapsed.
The four outgoing edges are ordered by priority, indicated by a grey number at
the beginning of an edge. For example, when a speech bubble appears and the
left key is pressed at the same time, the transition to the `text' state will
have a higher priority.
Finally, the edge (text, end) tests the removal of the speech bubble after 1
second. The state `end' is a stop state for all program models as this is the
game end.

The edge (start, text) is annotated with a timed condition `force at 31s' in
brackets. This is an abbreviated form to denote that the edge has to be taken
within 31 seconds after the program has started, and if it is not taken
by the program under test, then an error is reported. More generally, each edge
in our implementation can have a timed condition from game start denoted with
`force at' and a timed condition, described by `force after', indicating the
maximal time difference between edge transitions.
Note that the \Whisker testing framework allows users to accelerate program executions using a customisable acceleration factor. In such cases, timed conditions are automatically modified with respect to the chosen acceleration factor.

\Cref{fig:codeBowlWrong} shows an erroneous implementation produced by a
student, in which the bowl is 
missing
an action on right key presses, and the speech bubble at the end is shown for two
seconds instead of one.
The following failures are reported by the bowl model in \cref{fig:BowlProgramModel}:
\begin{itemize}
	\item Bowl.x+ missed: When the model is given a right key as input, it takes the transition with the `right key' condition. This expects an increase of Bowl.x as effect, which is not reported by the \Scratch VM. The model then reports this as a missing functionality of the program under test.
	\item No output of Bowl (End) after 1s: The edge (text, end) is triggered after 1s and checks whether the bowl still has a speech bubble containing `$/$End$/$'. The model reports the output is not removed after 1s.
\end{itemize}

\subsection{Apple Model}

\begin{figure}[tb]
	\centering
	\includegraphics[width=1\linewidth]{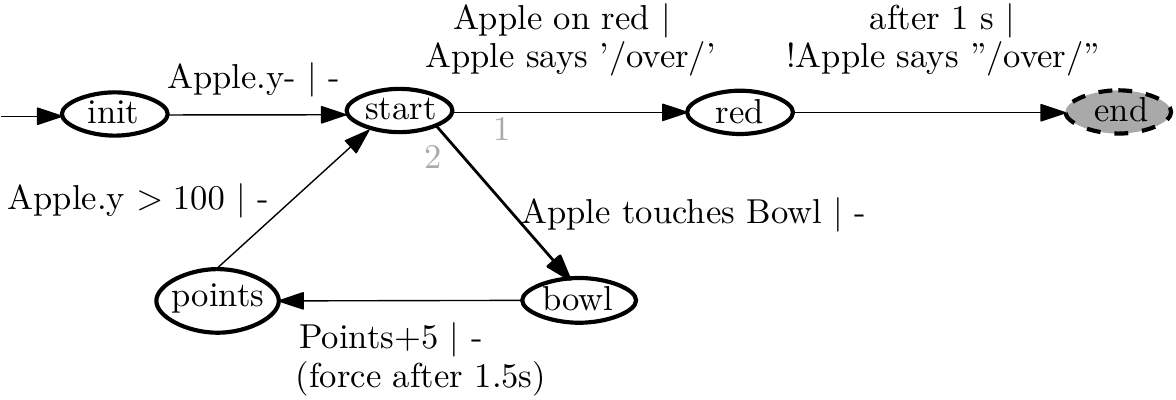}
	\caption{Model defining the apple's behaviour.}
	\label{fig:appleModel}
\end{figure}

The behaviour of the apple sprite is represented by \cref{fig:appleModel}.
Initially, the model expects the apple to be moved to the top of the screen and to start falling down.
The model then waits for the apple to touch either the bowl, or the red bar at
the bottom of the screen. 
If the bowl is touched, the model moves to the state
`bowl', which expects the points to increase by five within 1.5 seconds.
Hereafter, edge~(points, start) represents the expected placement of the apple to the top of the canvas by testing $y>100$, moving the apple up again for another round of the game as the specification states that the sprite should be set to
y~$=170$ if dropping into the bowl. 
%
Testing for the absolute value of y~$==170$ is disadvantageous on edge (points,~start) since a faulty implementation that sets the apple to value $y=150$ when touching the bowl, would lead to traversing both edges (start,~bowl) and (bowl,~points) and waiting for the condition of $y==170$ to be fulfilled, effectively halting the model in state `points'. 
Thus, to make the depicted model more robust against faulty implementations, we test against a value greater than 100 and move the check for the correct value of $y==170$ to another model.
When the apple hits the red line at the bottom of the
canvas a `Game~over' message should be displayed, represented by the regular
expression `$/$over$/$'. The model further expects this speech bubble to
disappear again after a second for a clean state at the end of the game.

\begin{figure}[tb]
	\centering
	\begin{subfigure}[b]{0.49\columnwidth}
		\centering
		\includegraphics[width=1\linewidth]{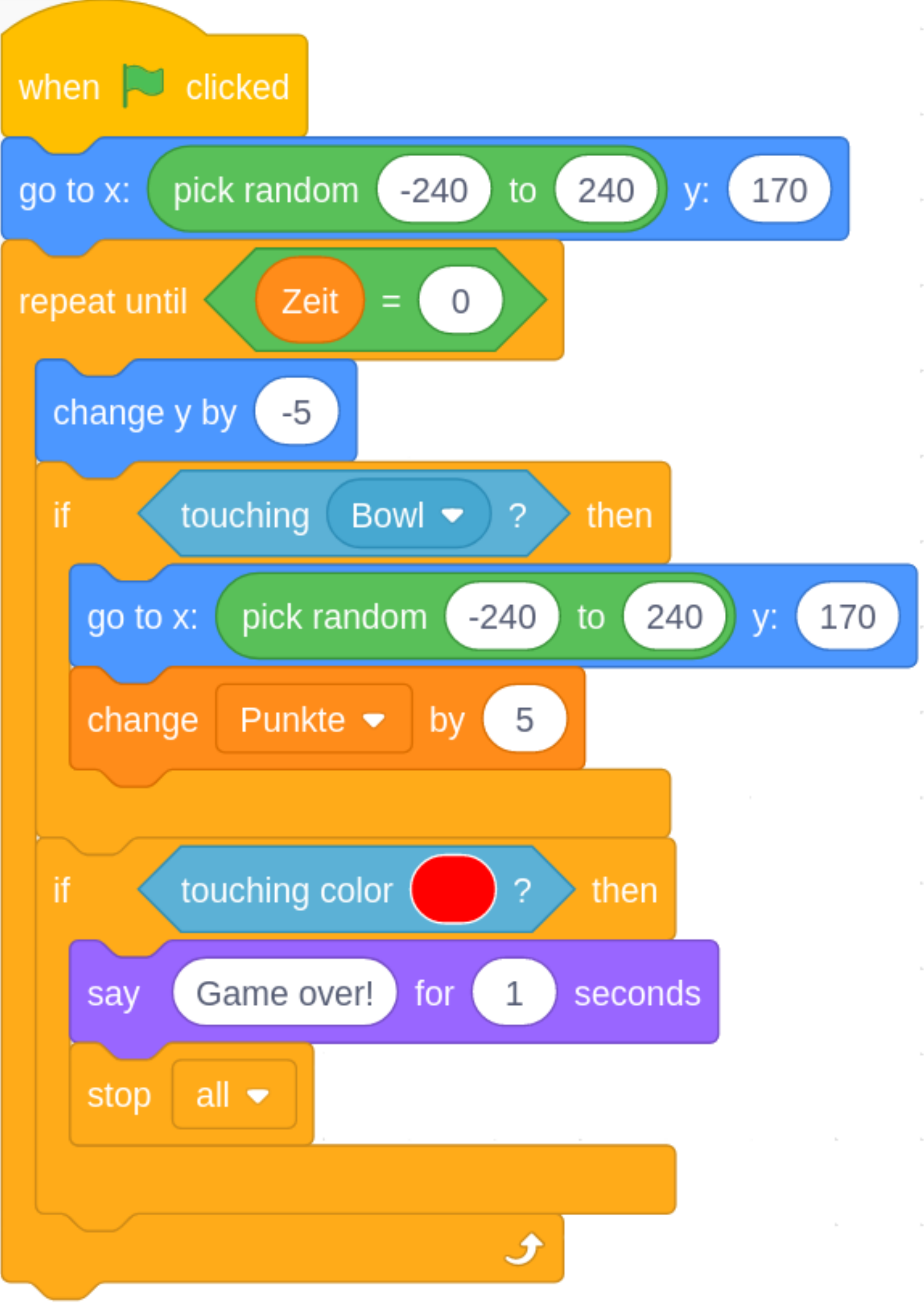}
		\caption{Sample solution for apple behaviour.}
		\label{fig:codeAppleCorrect}
	\end{subfigure}
	\hfill
	\begin{subfigure}[b]{0.49\columnwidth}
		\centering
		\includegraphics[width=1\linewidth]{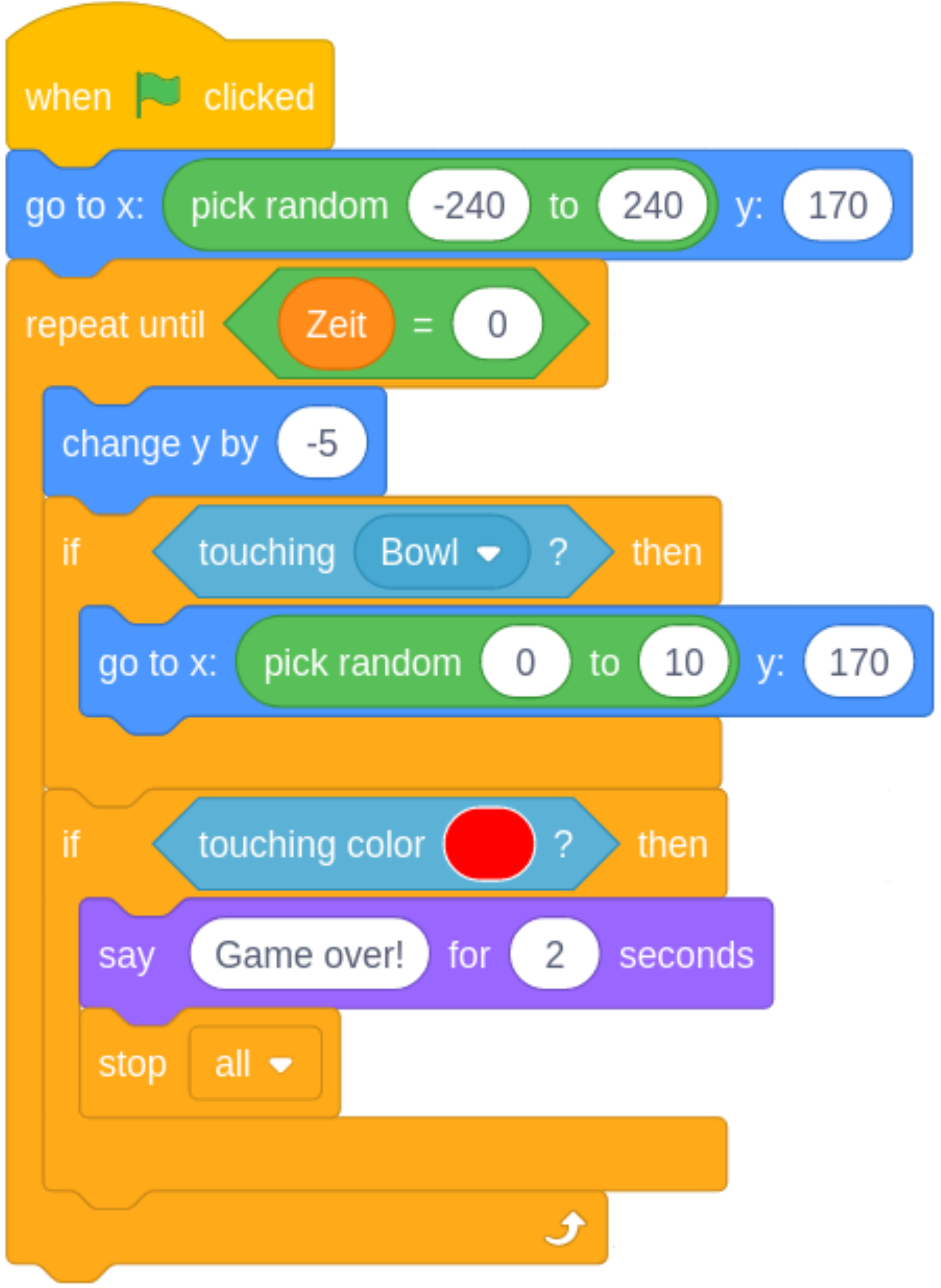}
		\caption{Code example with mistakes for apple sprite.}
		\label{fig:codeAppleWrong}
	\end{subfigure}
	\caption{Code examples for the apple sprite.}
	\label{fig:codeApple}
\end{figure}

Given a correct implementation of the apple's specifications in
\cref{fig:codeAppleCorrect} and a wrong implementation in
\cref{fig:codeAppleWrong}, where the increase in points is missing and the speech
bubble is shown too long, the apple model reports the following problems:
\begin{itemize}
	\item Points+5 missed: When the apple touches the bowl the model changes its state to `bowl'. After 1.5s the model reports that the transition to `points' has not been triggered and reports the condition as a failure. 
	\item No output of apple (over) after 1000ms: The edge~(red,~end) is triggered after 1s and checks whether the apple still has a speech bubble containing `$/$over$/$'. The model reports the output is not removed after 1s.
\end{itemize} 

\subsection{Bananas Model}
\begin{figure}[tb]
	\centering
	\includegraphics[width=0.9\linewidth]{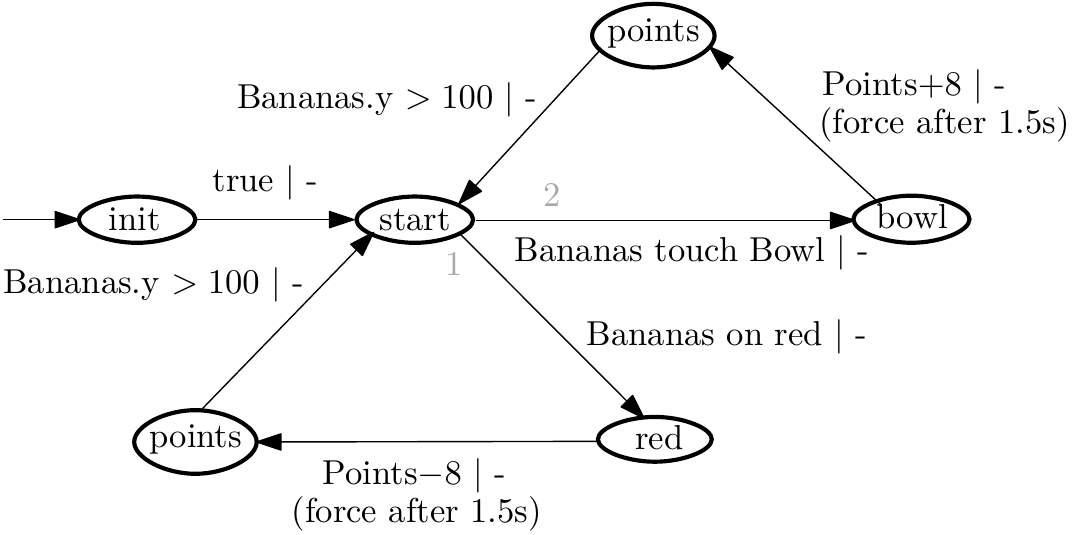}
	\caption{Model defining the bananas' behaviour.}
	\label{fig:bananaModels}
\end{figure}

For the bananas sprite the specification states that the player's points should be reduced by 8 when the bananas touch the red bar at the bottom and the game should resume by spawning the bananas again at the top of the canvas. When the bananas are caught by the bowl sprite the player's points are increased by~8. The model is shown in \cref{fig:bananaModels}. Again testing for a y value greater than 100 allows the continuous test of these models even if the fruit is not spawned to $y==170$ as specified.

\subsection{End Model}

\begin{figure}[tb]
	\centering
	\includegraphics[width=0.6\linewidth]{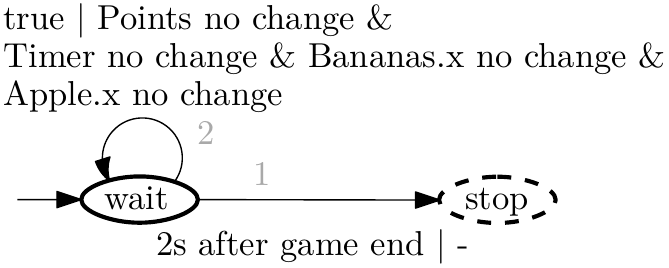}
	\caption{Model started after game end. Checks that no sprites are spawned or variables change.}
	\label{fig:end}
\end{figure}

End models make it possible to describe the behaviour at the end of program execution in a single model, rather than having to distribute it across multiple models.
For testing constraints after the fruit catching game has ended we implemented the end model shown in \cref{fig:end}. 
This model tests for two seconds that the sprite positions and the points and timer variables do not change, before ending the model test with the `stop' state. While the specification does not include any constraints for what should happen after a game ends, we found that this end model provides useful feedback for programs that do not halt. 

\subsection{User model}

\begin{figure}[tb]
	\centering
	\includegraphics[width=\linewidth]{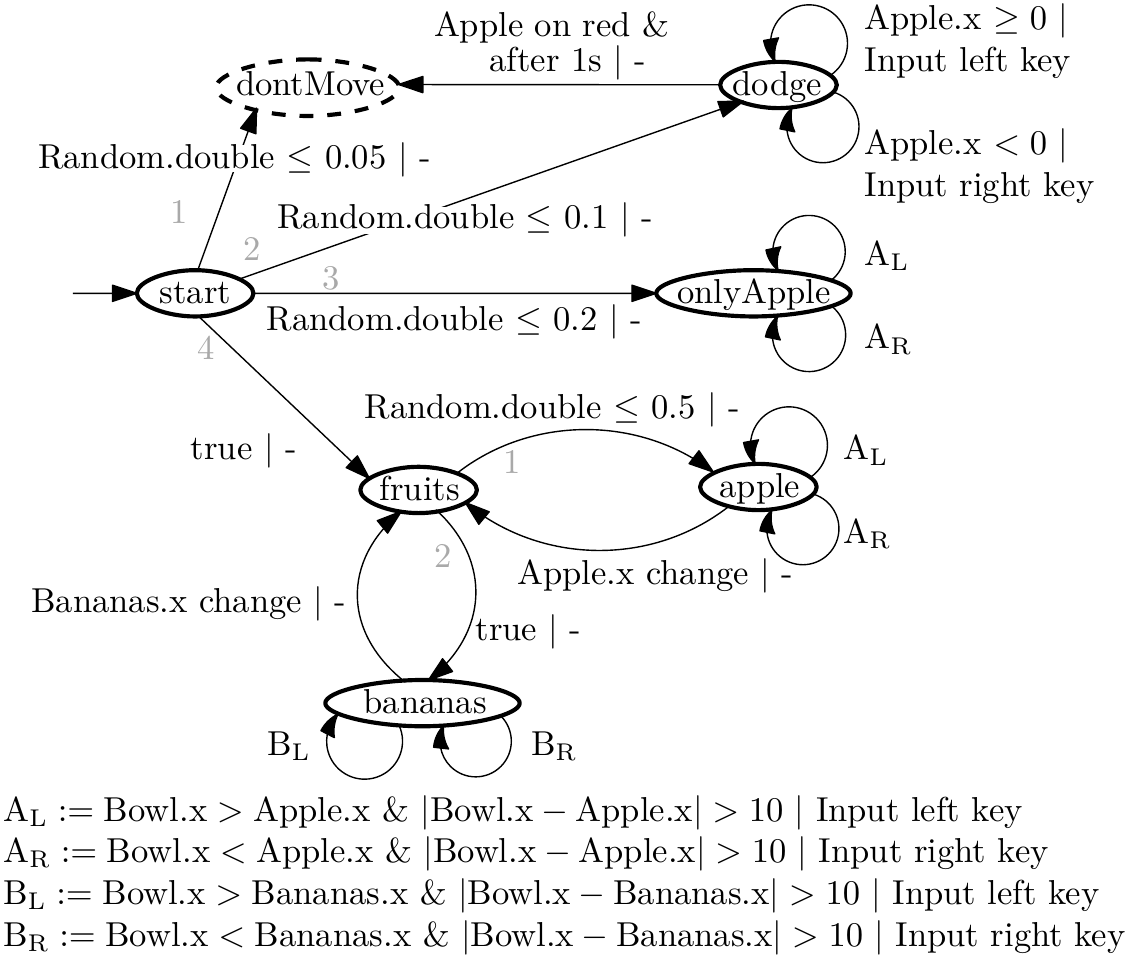}
	\caption{User model for the fruit catching game.}
	\label{fig:randomfruit-user}
\end{figure}

In order to fully automate the testing process, we also defined the user model shown in \cref{fig:randomfruit-user}, which generates inputs for the game based on randomness. For this purpose, we generate a random double in the range of $[0,1]$ and select the transition to take, including the corresponding input based on the randomly generated value.
From the `start' state, the model offers four types of
player behaviours: The model can choose to not move the bowl in state
`dontMove' at all, 
dodge the apple in state `dodge' by moving away from the apple to intentionally loose,
%
catch the apple continuously in state `onlyApple' to win or
decide randomly between the apple and bananas sprite in state `fruits'. When
deciding to catch a fruit the position of the bowl and the respective fruit are compared and a
self-looping edge inputs the correct key to move to the fruit. When the
difference in positions is smaller than ten it stops as the bowl should move in steps
of ten. Only when the fruit is caught and it respawns at the top, the next input is computed. We assume here that a respawn of a fruit sets the fruit to a new random x-position and therefore, react to it. The conditions y~$==170$ and y~$>100$ may either not be triggered or stop the movement towards the sprite too early.


\section{Evaluation}
To evaluate the effectiveness of model-based \Scratch testing we aim to answer the following research questions:
\begin{itemize}
\item \textbf{RQ1}: Are user models capable of exercising \Scratch programs thoroughly?
\item \textbf{RQ2}: Is model-based \Scratch testing capable of detecting faulty program states?
\end{itemize}
The proposed tool for testing \Scratch programs in a model-based manner is fully integrated into the \Whisker testing framework and available on GitHub\footnote{[October 2021] https://github.com/se2p/whisker}.

\subsection{Dataset}
The dataset used for evaluating the model-based \Scratch testing approach originates from a prior study~\cite{Whisker}, in which pupils visiting the sixth and seventh grades were taught programming in \Scratch.
As a final assignment at the end of the workshop, each student had to implement the fruit catching program on his/her own given a textual specification of the game.
To form a dataset, we first gathered every submitted student solution of the workshop which resulted in 41 fruit catching programs.
Then, similar to Stahlbauer et al.~\cite{Whisker} we removed three projects starting with a key press such as space instead of the greenflag event, because \Whisker assumes the projects to start on greenflag and the three projects thus would not react to any inputs.
In order to evaluate if the models meet the requirements defined within the textual specification, we also added the sample solution to the dataset.
Thus, altogether the dataset consists of 39 fruit catching versions, covering many different implementation approaches as the number of used \Scratch blocks ranges from 3 to~81.
Furthermore, in a manual grading process~\cite{keller2019improving}, a teacher assigned each student submission points in an interval of 0 to 30. 
The dataset contains poor student submissions with only 2 points but also perfectly working submissions with maximal points.
All in all, the dataset consists of 39 fruit catching programs sampled from a real-world application scenario, which reflects our targeted application scenario of helping a teacher in assessing many student solutions simultaneously.

\subsection{Methodology}
Model-based \Scratch testing aims to simplify the process of evaluating countless student submissions by offering a convenient alternative to writing error-prone JavaScript tests. 
Teachers can build visual models of their desired program behaviour, which are less susceptible to implementation errors and do not require users to be familiar with the JavaScript language.
When providing additional user models the testing process can be made independent of JavaScript tests.

For our research questions we compare the manually written JavaScript test suite by Stahlbauer et al.~\cite{Whisker}, which contains fixed user inputs  as well as intricate JavaScript code to react to various game situations, with the models described in \cref{sec:CaseStudy}. Two experiment settings are needed for this evaluation: The first one excludes our user model and only uses the 19 program models and one end model for the test runs while obtaining the user inputs from the JavaScript test suite. The other experiment is independent of the test suite and provides inputs based on our user model in \cref{fig:randomfruit-user}. Both test suite and models are based on the same textual specification and sample solution.

To minimize the influence of randomised behaviour in the fruit catching game or in the inputs, we repeat each experiment 30 times, where each run seeds the random number generator differently by \Whisker. 
When executing one run on a student implementation with model-based inputs we repeat the input generation 20 times to get different input sequences.
The same seeds are used for the different experiment settings. The \Whisker testing framework offers the option to accelerate test execution by a user-defined acceleration factor.
With the aim of decreasing the time required for evaluating all projects of the dataset, we speed up test execution by a factor of 10. 

\vspace{0.1cm}
\noindent\textbf{RQ1:} To evaluate how well model-based input generation can test a \Scratch program we compare the achieved block coverages of model-generated inputs against the inputs originating from the manually written test suite.
We report the achieved mean coverage values of both approaches together with the \emph{Vargha and Delaney} (A12) effect size for every project of the dataset.
We consider a result to be statistically significant if a  \emph{Mann-Whitney-U-test} reports a $p$-value smaller than 0.05.

\vspace{0.1cm}
\noindent\textbf{RQ2:} For model-based testing in \Scratch to be a valid alternative approach its test runs have to be able to detect faulty program state transitions as reliable as the manually written JavaScript test suite. For this, we compare the number of erroneous program state transitions found by our defined models against the test suite.
To prevent the used method of input generation to influence the results, we performed the experiment once with model input generation and once with inputs from the test suite.

\subsection{Threats to Validity}
\noindent \textbf{External Validity:} Our dataset, composed of 39 fruit catching programs, contains implementations that are very similar to the sample solution as well as fruit catching versions that deviate significantly from the desired program behaviour.
Additionally, the grades manually awarded by a teacher, ranging from 2 to 30 points, indicate that the dataset contains many different program behaviours, of which some are entirely correct and others mostly faulty.
However, due to the lack of data originating from similar application scenarios, the experiments were only conducted on various fruit catching versions and thus might not generalise well to other programs.

\vspace{0.1cm}

\noindent \textbf{Internal Validity:} Because the fruit catching game as well as the inputs derived from the user model include randomised behaviour, 
we repeat each experiment 30 times and report the achieved averages and A12 effect sizes across all repetitions.
Another threat to internal validity is the use of manually developed models and test cases, as already minor changes in both may already lead to significant variations in the experiment results. The results of experiments with model-based inputs also depends on the quality of the user model. Because our user model is based on randomness, we chose 20 repetitions to create different input sequences, which could vary with different probabilities on the transitions and less repetitions.

\vspace{0.1cm}

\noindent \textbf{Construct Validity:} The achieved block coverage, a measure similar to statement coverage in mature programming languages, is used to determine how well a given approach can exercise \Scratch programs.
However, because most program states are often already covered by simply starting a given \Scratch program, block coverage may not always be a good indicator of how thoroughly a given approach explores a student submission.
Furthermore, we compare the effectiveness of the model-based approach in finding faulty program behaviours against a manually written test suite by counting the number of reported erroneous program states.
Even though the models and the manually written test suite were developed with the intention to reflect the textual specification of the fruit catching game, a human evaluator may not always agree that reported program states are indeed faulty.

\subsection{RQ1: Are User Models Capable of Exercising \Scratch Programs Thoroughly?}
\begin{table}[tb]
    \centering
    \caption{Mean coverage and A12 effect size: UserModel-Inputs vs. TestSuite-Inputs. Values in boldface indicate strong statistical significance with $p <$ 0.05.}
\vspace{-0.7em}
    \label{tab:RQ1}
    \resizebox{\columnwidth}{!}{
    \begin{tabular}{lrrrlrrr}
        \toprule
        Project & \rotatebox{90}{UserModel} & \rotatebox{90}{TestSuite} & \rotatebox{90}{A12} &
        Project & \rotatebox{90}{UserModel} & \rotatebox{90}{TestSuite} & \rotatebox{90}{A12}\\
        \midrule
        C6\_01 & 70& 68 & \textbf{0.65} &  C7\_03 &57 & 57 & 0.5  \\
        C6\_02 & 73 & 73 & 0.5 & C7\_04 & 74 & 73 & 0.53 \\
        C6\_03 & 94 & 94 & 0.5 & C7\_05 & 88 & 82 & \textbf{1.0} \\
        C6\_04 & 97 & 94 & \textbf{0.88} & C7\_06 & 77 & 75 & \textbf{0.75}  \\
        C6\_05 & 87 & 88 & 0.47 & C7\_07 & 87 & 87 & 0.5 \\
        C6\_06 & 32 & 35 & \textbf{0.42} & C7\_08 & 95 & 89 & \textbf{0.98} \\
        C6\_07 & 94 & 98 & \textbf{0.0} & C7\_09 & 71 & 71 & 0.5 \\
        C6\_08 & 80 & 80 & 0.5 & C7\_10 & 99 & 98 & \textbf{0.97} \\
        C6\_09 & 100 & 100 & 0.5 & C7\_11 & 99 & 100 & 0.48 \\
        C6\_10 & 33 & 33 & 0.5 & C7\_12 & 100 & 100 & 0.5 \\
        C6\_11 & 98 & 97 & \textbf{0.74} & C7\_13 & 88 & 84 & \textbf{1.0} \\
        C6\_12 & 97 & 97 & 0.5 & C7\_14 & 99 & 99 & 0.5 \\
        C6\_13 & 69 & 69 & 0.5 & C7\_15 & 95 & 90 & \textbf{1.0} \\
        C6\_14 & 80 & 80 & 0.5 & C7\_16 & 91 & 90 & 0.55 \\
        C6\_15 & 85 & 89 & \textbf{0.0} & C7\_17 & 93 & 94 & 0.45 \\
        C6\_16 & 100 & 98 & \textbf{1.0} & C7\_18 & 52 & 52 & 0.42\\
        C6\_17 & 100 & 100 & 0.5 & C7\_19 & 97 & 93 & \textbf{0.97} \\
        C6\_18 & 94 & 91 & \textbf{1.0} & C7\_20 & 67 & 50 & \textbf{1.0} \\
        C7\_01 & 74 & 65 & \textbf{1.0} & Sample & 100 & 100 & 0.5 \\
        C7\_02 & 92 & 91 & \textbf{0.58} & Summary & 84.27 & 82.87 & 17/\textbf{15}/39\\
        \bottomrule
    \end{tabular}
    }
\vspace{-1em}
\end{table}

User models offer a less error-prone alternative to explicitly specifying \Scratch inputs within a test suite written in JavaScript. 
However, those user models are only of use if they are able to exercise \Scratch programs as thoroughly as manually written tests.

For each project of the dataset and across all 30 experiment repetitions, \cref{tab:RQ1} compares the achieved block coverages of inputs derived from a user model against inputs obtained by the test suite.
Furthermore, we report the A12 effect size for each project individually and highlight statistically significant results with a $p$-value $<$ 0.05 in boldface.
Every project of the dataset, except the sample solution, starts with the letter \emph{C} followed by a number, which encodes the class of the student who created the given project.

Because the test suite and the user model were implemented to reflect the desired program behaviour defined within the textual specification, both approaches reach full project coverage on the provided sample solution. 
With a mean block coverage of 84\%, model-generated inputs covered slightly more blocks than the manually written test suite. 
Regarding the A12 effect size, in 17/38 student submissions, with 15 of them being statistically significant results, the user model benefits from its flexible input generation approach and reaches higher coverage values than the static test suite, which is specifically tailored to the control flow of the sample solution.
Furthermore, in 14/38 student projects, the user model and the test suite achieve the exact same coverage.
This may indicate that the program coverage may not be improvable when changing input generation techniques. The fact that the program coverage is also influenced by dead code blocks, unreachable code and wrong thread starting hats supports this. For example, in \emph{C6\_S08} we found dead code blocks without connected thread-starting hats that will never be executed independent from the input. Other implementations had unreachable code based on semantics e.g. putting the apple on red in the initial step, stopping with the testing on red in the second step and not being able to go into other code. 
Without such problematic scenarios the program coverage would be higher. 

Some projects have smaller project coverage when using model-based inputs.
Projects C6\_06 and C6\_15 are setting their fruit sprites at fixed positions with at least one in the middle within range of the bowl without movement. 
As the user model is flexible and tests the distance of bowl to fruit before providing any input, the model does only generate either left or right key and misses the other one.
The code to the missing key is therefore never triggered and the code coverage is lower. 
%
%
Another difference between the inputs generated is that the model can choose to catch bananas, while the test suite only concentrates on the apple.
In student submission C6\_05 we have the problem that catching the bananas effectively halts the program, again only producing left inputs. 

Projects C6\_07 and C7\_17 do not stop when the apple touches red and also do not move the apple up when touching the bowl. The model test stops shortly after the first apple touches red, never playing for the entire duration of 30 seconds. The code coverage of the model inputs therefore never reaches the code for event `timer $==0$' and has lower coverage for these projects. This could be easily prevented by changing the end model to test for at least 32 seconds.

\vspace{1em}
 \begin{tcolorbox}
	\textbf{Summary RQ1}: Since the user model performs equally well or even better than a manually written test suite in 31/38 student submissions, inputs derived from a model pose a powerful alternative to error-prone tests encoded in JavaScript.
 \end{tcolorbox}

\subsection{RQ2: Is Model-Based \Scratch Testing Capable of Detecting Faulty Program States?}
Similar to the process of generating inputs, model-based \Scratch testing offers an alternative to manually written test suites for evaluating the functional correctness of a given \Scratch program.
However, to be of any use, models must be capable of detecting faulty program states reliably.
Thus, to explore the effectiveness of model-based \Scratch testing, we compare the number of found program failures of the model-based approach against the failed test cases in the manually written test suite.

\begin{figure}[t]
	\centering
	\includegraphics[width=\columnwidth]{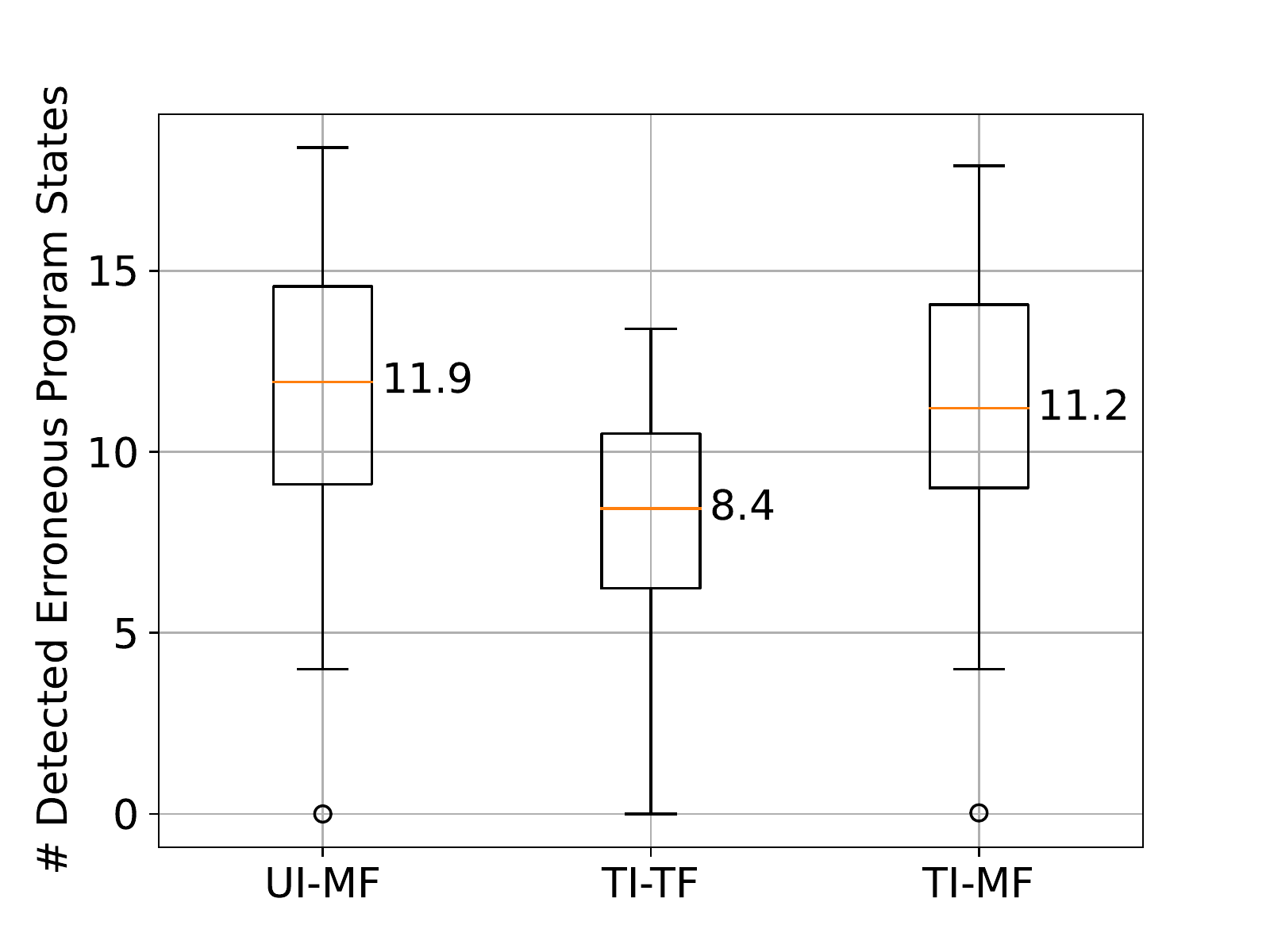}
	\caption{Number of detected erroneous program states: UserModelInput-ModelFailures (UI-MF) vs. TestSuiteInput-TestFailures (TI-TF) vs. TestSuiteInput-ModelFailures (TI-MF).}
	\label{fig:RQ2}
\end{figure}

\cref{fig:RQ2} shows the number of detected program states that deviate from the textual specification across all projects and experiment repetitions.
Note that every approach contains at least one project in which not a single faulty program state was detected.
This project corresponds to the sample solution and indicates that neither the test suite nor the models tend to generate false positives.
Overall, the combination of user model input generation and model-based testing (UI-MF) detects the most failures with a median of 11.9 reported program errors.
Furthermore, since both model-based approaches (UI-MF \& TI-MF) significantly outperform the manually written test suite (TI-TF), the results show strong evidence that model-based \Scratch testing is indeed capable of finding far more faulty program behaviours than manually written tests, regardless of how test inputs were generated.
%

\vspace{1em}
 \begin{tcolorbox}
	\textbf{Summary RQ2}: Model-based \Scratch testing is a convenient  alternative to manually written JavaScript tests and can detect far more faulty program states.
 \end{tcolorbox}


\section{Conclusions}\label{Conclusions}

The game-like and randomised nature of visual programming languages can make testing challenging. 
In this paper we proposed a model-based approach for testing \Scratch programs in the \Whisker framework. 
Together with model-based input generation this provides a fully automated testing approach, the effectiveness of which was indicated by our experiments.

The \Whisker tool contains an integrated GUI to create models for \Scratch
programs, and to apply model-based testing to programs under test. Our hope is
that a convenient tool will make building models easy and approachable for the
\Scratch target audience, such as teachers and content providers creating
tutorials. We provide \Whisker as an open source tool at the following URL:
\begin{center}
	\url{https://github.com/se2p/whisker}
\end{center}

While even simple models can quickly reveal faults in implementations of
\Scratch programs, our case study shows that creating a faithful representation
of a specification can be non-trivial. A central aspect of our modelling approach is therefore to allow the creation of many small models, rather than a single big model. However, the large number of 21 models of our case study program with a very detailed specification suggest that further support may be necessary for users. In particular, given that an example solution is usually available in our application scenario, model inference seems like a promising direction for future research.
Users could also be supported by techniques like graph minimisation to provide further insights into eliminating erroneous state transitions. Another promising avenue of further research is using the models to automatically derive hints and feedback for learners.

\section*{Acknowledgements}
This work is supported by DFG project
FR 2955/3-1  ``Testing, Debugging, and Repairing
Blocks-based Programs''.

\balance
\bibliographystyle{IEEEtran}
\bibliography{references}

\end{document}